\newcommand{\be}{\begin{equation} }
\newcommand{\ee}{\end{equation} }
\newcommand{\bea}{\begin{eqnarray} }
\newcommand{\eea}{\end{eqnarray} }

\newcommand{\PHCX}{(C$_4$H$_{12}$N$_2$)Cu$_2$(Cl$_{1-x}$Br$_x$)$_6$\xspace}

\newcommand{\TCCL}{TlCuCl$_3$\xspace}
\newcommand{\TCCX}{ Tl$_{1-x}$K$_x$CuCl$_3$\xspace}

\newcommand{\IPA}{IPA-CuCl$_3$\xspace}
\newcommand{\IPAXf}{(CH$_3$)$_2$CHNH$_3$Cu(Cl$_{1-x}$Br$_x$)$_3$\xspace}
\newcommand{\IPAX}{IPAX\xspace}
\newcommand{\IPAXP}{IPAX-0.05\xspace}

\newcommand{\DTNX}{{NiCl$_{2-2x}$Br$_{2x}\cdot$4SC(NH$_2$)$_2$}\xspace}

\documentclass[ps,prb,reprint,showpacs,longbibliography,superscriptaddress]{revtex4-1}

\usepackage{amsmath,amssymb,bm}
\usepackage{graphicx}
\usepackage{color}
\usepackage{xspace}

\begin{document}

\title{Field-induced ordering in a random-bond quantum  spin ladder compound with weak anisotropy.}

\author{G.~S.~Perren}
\affiliation{Laboratory for Solid State Physics, ETH Z\"urich, 8093 Z\"urich, Switzerland.}

\author{W.~E.~A.~Lorenz}
\affiliation{Laboratory for Solid State Physics, ETH Z\"urich, 8093 Z\"urich, Switzerland.}

\author{E.~Ressouche}
\affiliation{INAC, SPSMS, CEA Grenoble, 38054 Grenoble, France}

\author{A. Zheludev}
\email{zhelud@ethz.ch}
\homepage{http://www.neutron.ethz.ch/}
\affiliation{Laboratory for Solid State Physics, ETH Z\"urich, 8093 Z\"urich, Switzerland.}

\date{\today}

\begin{abstract}
The field induced quantum phase transitions in the disorder-free and disordered samples of the spin ladder material \IPAXf are studied using magnetic calorimetry and magnetic neutron diffraction on single crystal samples. Drastically different critical indexes and correlation lengths in the high field phase are found for two different orientations of the applied field. It is argued that for a field applied along the crystallographic $c$ axis, as in previous studies,\cite{Hong2010PRBRC} the transition is best described as an Ising transition in random field and anisotropy, rather than as a magnetic Bose Glass to Bose Condensate transition.

\end{abstract}


\maketitle
\section{Introduction}
Shortly after the realization that field-induced magnetic ordering in gapped quantum paramagnets can be viewed as a Bose-Einstein condensation (BEC) of magnons,\cite{Giamarchi1999,Rice2002,Giamarchi2008} a new research thrust was aimed at understanding the corresponding transition in disordered spin systems (for a review see Ref.~\onlinecite{Zheludev2013}). The idea is that in the presence of disorder, the magnetic BEC state may be preceded by the magnetic analog of the long sought for Bose Glass (BG) state.\cite{Fisher1989,Yu2010,Yu2012,Zheludev2013} The organic quantum spin ladder compound \IPAXf (\IPAX for short) was one of the first materials studied in this context.\cite{Manaka2008,Manaka2009,Hong2010PRBRC} Indeed, neutron scattering experiments under applied field revealed a BG-like gapless, magnetizable (``compressible'') yet disordered magnetic state.\cite{Hong2010PRBRC} Most of the experimental work since was done on other disordered quantum magnets, such as \DTNX\cite{Yu2012,Wulf2013} and \PHCX\cite{Huevonen2012,Huevonen2012-2,Huevonen2013}. The main effort was aimed at understanding the critical exponents of the quantum phase transition between the magnetic BEC and BG phases. Of particular interest is the experimentally accessible exponent $\phi$, which defines the field-temperature phase boundary: $T_c=(H_c-H)^\phi$. The original work of Fisher et al. \cite{Fisher1989} predicted $\phi>2$, as opposed to $\phi=2/3$ in the absence of disorder. This theoretical result was later challenged by  QMC calculations that suggested $\phi\approx 1.1$,\cite{Yu2012} but were in turn contradicted by a later study consistent with Fischer's original arguments.\cite{Yao2014} So far, experiments on materials like \DTNX \cite{Yu2012,Wulf2013,Wulf2015} and \TCCX\cite{Yamada2011,Zheludev2011} have only added to the controversy, and the issue remains unresolved.

To date, there have been no systematic studies of critical exponents in the ``original'' BG-prototype compound \IPAX. Moreover, from the start it was clear that the field-induced ordering transition in this material is not exactly a BG to BEC one. In \IPAX the high field phase has static, but only short range magnetic order, with a history dependent correlation length,\cite{Hong2010PRBRC} in contrast to true long range order expected for a magnetic BEC. A convincing explanation for this behavior is lacking. In the present work we address these issues in a series of bulk measurements and additional neutron diffraction studies on the $x=0.05$ and $x=0$ systems. We show that the previously observed short-range order is directly linked to a weak magnetic anisotropy present already in the parent compound \IPA. We further argue that in the context of \IPAX, and perhaps in most other quantum paramagnets with chemical disorder,  it is more appropriate to speak of spin freezing in a random field environment, rather than of magnon condensation in a magnetic BG.

\IPA  crystallizes in a triclinic space group $P\overline{1}$ with $a=7.78$~\AA, $b=9.71$~\AA, $c=6.08$~\AA, $\alpha=97.26^\circ$, $\beta=101.05^\circ$, and $\gamma=67.28^\circ$.\cite{Manaka1997,Manaka1998} The magnetic properties are due to $S=1/2$ Cu$^{2+}$ ions that form weakly coupled two-leg ladders directed along the crystallographic $a$ axis.\cite{Masuda2006} The ground state is a spin singlet. The magnetic exitation spectrum features a gap of $\Delta=1.2$ meV and dispersive triplet excitations characteristic of a the spin ladder model.\cite{Masuda2006,Zheludev2007} Based on these neutron data, numerical modeling allowed to determine the relevant exchange constants.\cite{Fischer2011} The magnon triplet is split into a singlet and doublet by 0.05~meV due to a very weak easy-axis anisotropy of exchange interactions.\cite{Manaka2001,Nafradi2011} The easy axis is roughly along the crystallographic $b$ direction. The anisotropy is also manifest in the gyromagnetic tensor with $g_a=2.05$, $g_b=2.22$, and $g_c=2.11$.\cite{Manaka2007-2} The field-induced ordering transition occurs in $H_c\sim9.7$~T.\cite{Manaka1998} The transition can to some extent be viewed as a BEC of magnons.\cite{Garlea2007} This picture is disrupted by the residual anisotropy which leads to are-opening of a small gap in the ordered state.\cite{Zhao2015} In the Br-substituted compound \IPAX the crystal structure is almost unchanged up to $x=0.13$.\cite{Manaka2001} The halogen substitution does not directly affect the magnetic Cu$^{2+}$ ions, but instead randomizes the strengths of magnetic interactions. As mentioned above, the result is a Bose-glass like precursor to the field-induced ordered phase.\cite{Manaka2008,Manaka2009,Hong2010PRBRC}

\section{Experimental details}
The present study was performed on the same fully deuterated \IPA and $x=0.05$ \IPAX (\IPAXP for short) single crystal samples as used in Refs.~\onlinecite{Garlea2007} and \onlinecite{Hong2010PRBRC}, respectively. Thermal-relaxation calorimetry was carried out on $\sim1$~mg fragments, using a Quantum Design 14~T PPMS with the $^3$He-$^4$He dilution refrigerator insert. The data were collected in two field orientations. A close to axial geometry was realized with the magnetic fields $\mathbf{H}$ applied  along the $\mathbf{b}^*$-axis, which is conveniently perpendicular to a natural cleavage plane of the crystals. In this orientation the field is only $\sim 15^\circ$ misaligned with the magnetic easy axis.\cite{Manaka2007-2} Alternatively, the field was applied along the crystallographic $\mathbf{c}$-direction, making it almost perpendicular to the anisotropy axis (transverse geometry).

A new series of neutron diffraction measurements were carried out on the $x=0.05$ \IPAX compound  using fully a deuterated 150~mg single crystal. The magnetic field was applied along the $\mathbf{b}$ direction. This setting almost exactly realizes the axial geometry, in contrast to previous studies of Ref.~\onlinecite{Hong2010PRBRC} where the field was applied in the transverse direction. In our experiments, the sample environment was a 12~T cryomagnet with a dilution refrigerator insert.  Diffraction was carried out using  $\lambda=1.218$~\AA\~ neutrons from a Cu-$(200)$ monochromator.

\section{Results and data analysis}
\subsection{Calorimetry}
\begin{figure}[!htbp]
	\includegraphics[width=0.8\columnwidth]{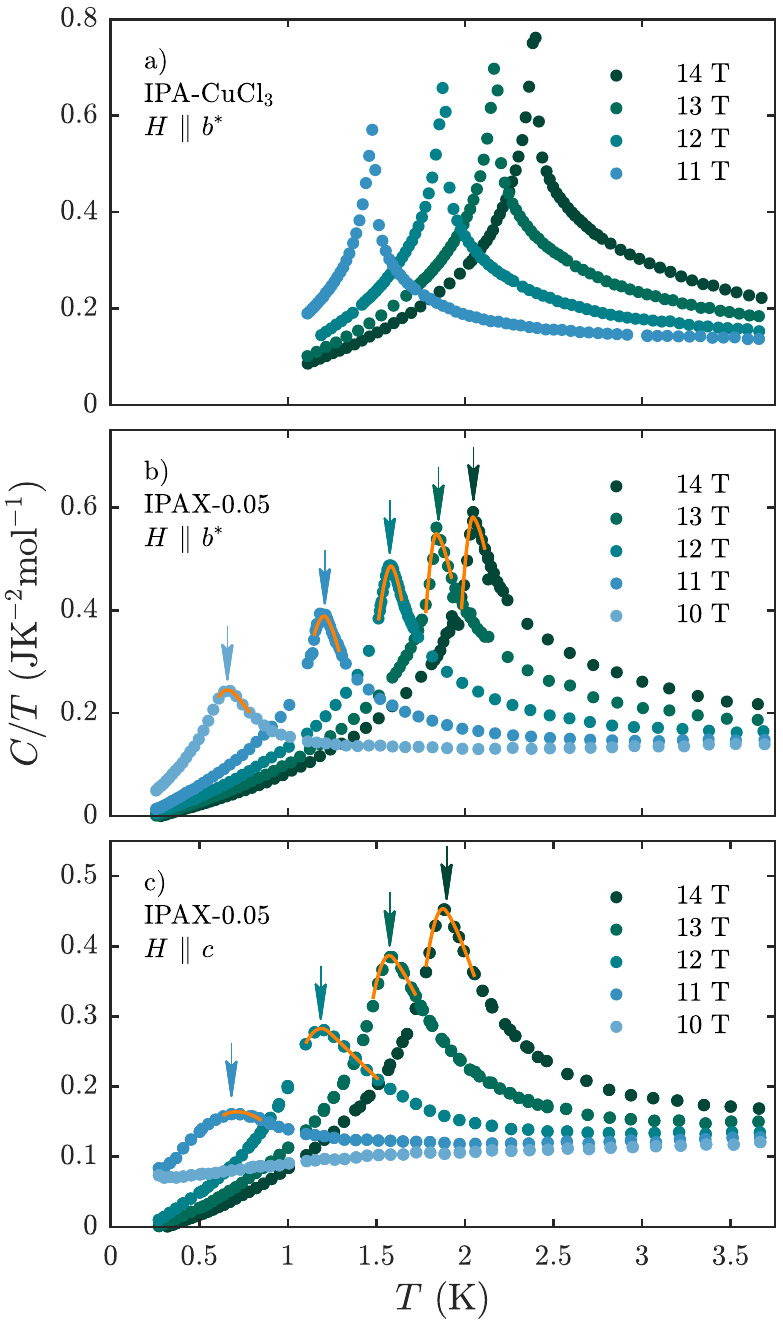}
	\caption{\label{CvsT} Magnetic specific heat vs. temperature measured in \IPA (a) and \IPAX (b) for a field applied almost parallel to the anisotropy axis, and \IPAX in a field almost perpendicular to it (c). Solid lines are empirical fits to locate the maximum.}
\end{figure}

\begin{figure}[!htbp]
	\includegraphics[width=0.8\columnwidth]{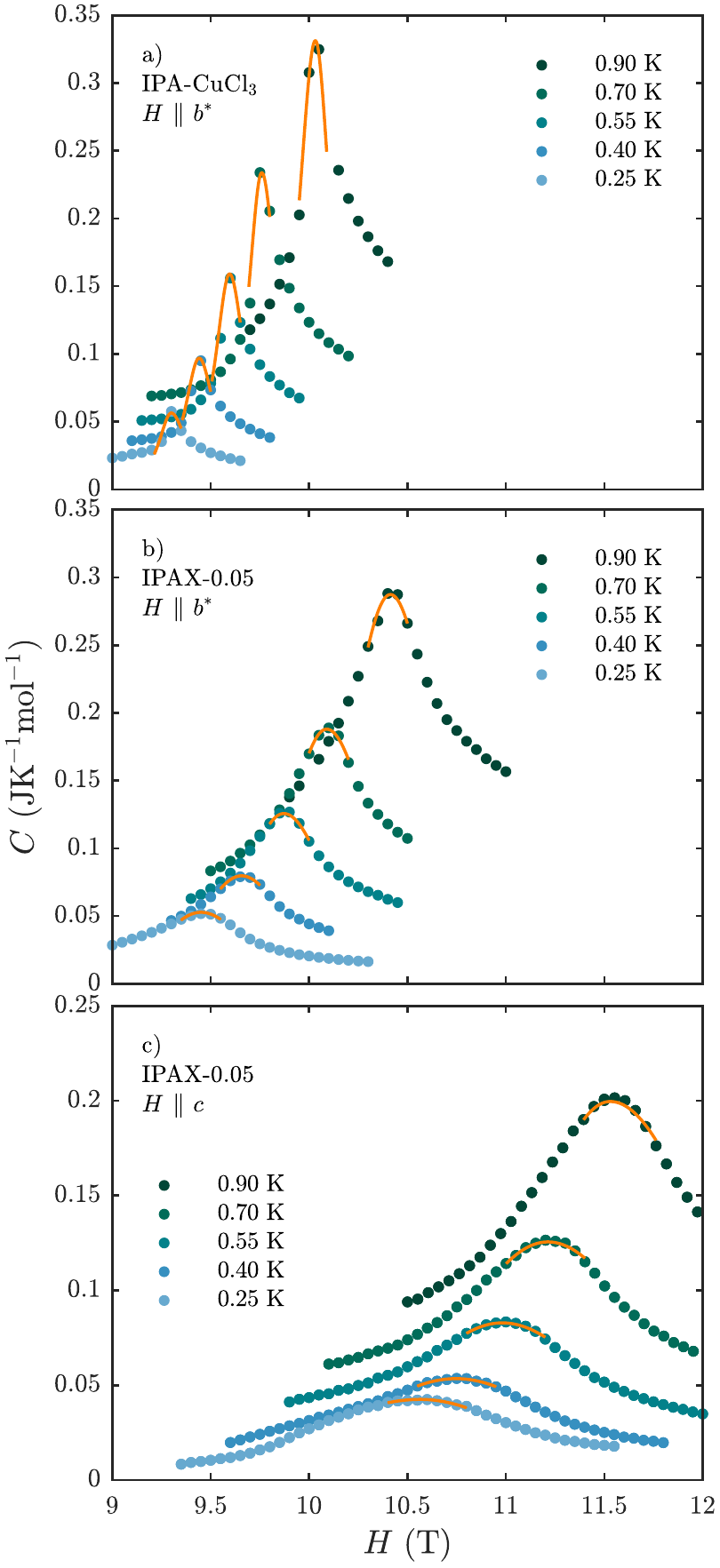}
	\caption{\label{CvsH} Specific heat vs. applied field, as measured in \IPA (a) and \IPAX (b,c). The field is almost parallel to the anisotropy axis in (a,b), and almost perpendicular to it în (c). Solid lines are empirical fits to locate the maximum.}
\end{figure}

Typical magnetic $C(T)$ and  $C(H)$ curves measured for \IPAX and \IPA are shown in Figs.~\ref{CvsT} and~\ref{CvsH}, respectively. In the constant-field data, the Debye lattice contribution was measured in zero applied field and subtracted from the data shown. In all cases, a peak in the specific heat curve was taken as a signature of the onset of long range order. For the disorder-free sample the lambda anomalies are perfectly sharp, but somewhat rounded in \IPAXP. In that case, the lambda anomaly is particularly broad in the transverse-field $\mathbf{H}\| \mathbf{c}$ geometry, especially below $T\sim 1$~K.

\begin{figure}[!htbp]
	\includegraphics[width=\columnwidth]{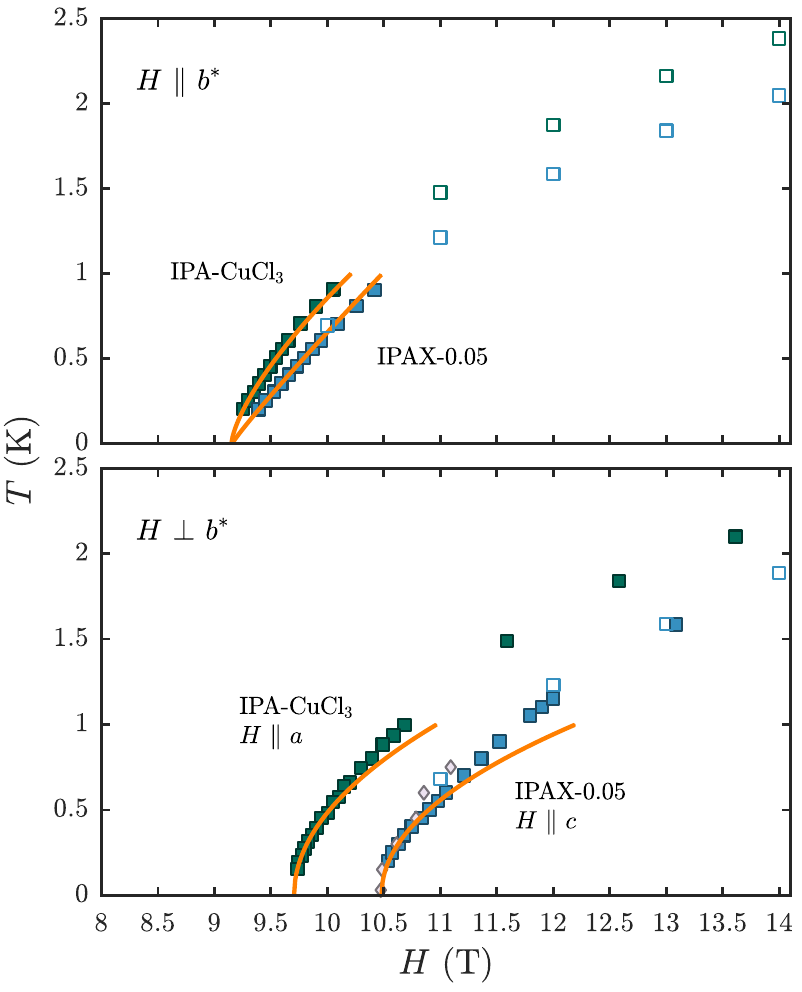}
	\caption{\label{phase} Magnetic phase diagrams measured for \IPA (top) and \IPAXP (bottom) for magnetic fields applied  almost parallel to the easy anisotropy axis ($\mathbf{b}^\ast$-direction), and almost transverse to it. Solid and open squares are positions of specific heat maxima in constant-temperature and constant-field scans, respectively. Diamonds are neutron diffraction data from Ref.~\onlinecite{Hong2010PRBRC}. The $\mathbf{H}\bot\mathbf{b}^\ast$ data for the parent compound are from Ref.~\onlinecite{Tsujii2009}. Solid lines are power law fits as described in the text. }
\end{figure}

To reconstruct the corresponding phase diagrams, the calorimetry data were analyzed with empirical ``broadened power-law'' fits for constant-field scans and ``split-Lorentzian'' fits for constant-$T$ scans,\cite{Huevonen2012} shown in solid lines in Figs.~\ref{CvsT} and~\ref{CvsH}, respectively. For the disorder-free system the transition was simply identified with the sharp maximum in the measured data. The results are shown in Fig.~\ref{phase}. The \IPAXP phase boundary data measured in the transverse geometry using neutron diffraction in Ref.~\onlinecite{Hong2010PRBRC} are also shown. They appear to be consistent with the present calorimetric measurements, despite the larger scattering of data points.
Also shown in Fig.~\ref{phase} are data for the parent compound from Ref.~\onlinecite{Tsujii2009} collected in the transverse-field geometry $\mathbf{H}\|\mathbf{a}$.

\begin{figure}[!htbp]
	\includegraphics[width=\columnwidth]{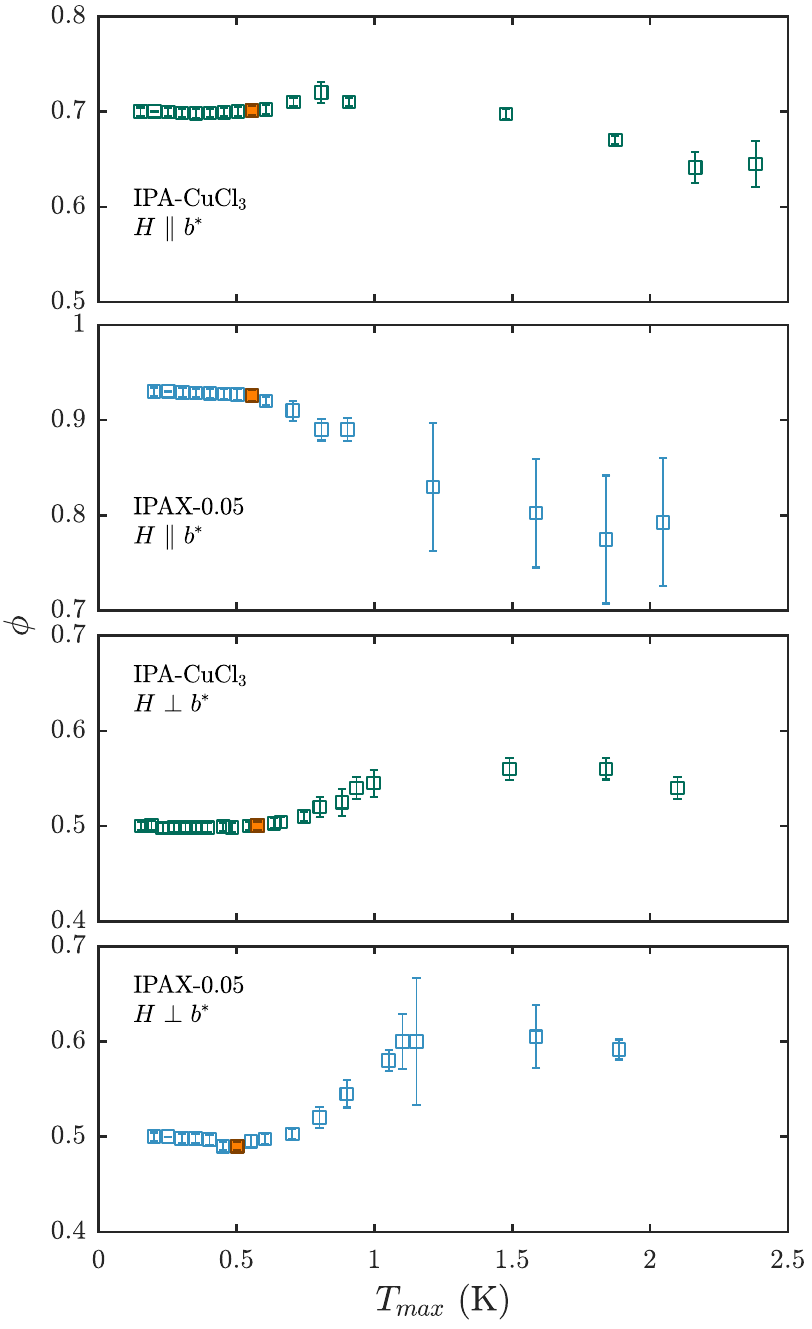}
	\caption{\label{phi} Shrinking-fit-window analysis of the measured phase boundaries. The plots show the least-squares fitted values of the phase boundary exponent $\phi$ vs. the temperature range used for the fit. The other two fit parameters, $H_c$ and an overall scaling factor, are not shown. }
\end{figure}

In order to estimate the exponents $\phi$, the $C(H)$ phase-boundary data were analyzed using power-law fits. The latter were performed in a progressively shrinking temperature range, following the procedure described in Ref.~\onlinecite{Huevonen2012}. The fit parameters were $\phi$, $H_c$ and an overall scale factor. The dependencies of the fitted value of $\phi$ on the temperature range used is plotted in  Fig.~\ref{phi}. We see that the results are very stable for all fitting ranges below $T_\mathrm{max}=0.6$~K.  The key parameters obtained in  that temperature range are summarized in Table~\ref{phih}. The corresponding fits are shown in solid lines in Fig.~\ref{phase}.

\begin{table}[!htpb]
	\begin{ruledtabular}
	\begin{tabular}{lcc}
		    & $\phi$ & $H_c$ (T)  \\ \hline
$\mathbf{H}\|\mathbf{b}^\ast$:\\
		\IPA  & 0.71(1)              & 9.15(1)                \\
		\IPAXP & 0.93(1)              & 9.16(1)              \\ \hline

$\mathbf{H}\bot\mathbf{b}^\ast$:\\
		\IPA ($\mathbf{H}\|\mathbf{c}$)  & 0.50(1)              & 9.99(1)              \\
		\IPAXP ($\mathbf{H}\|\mathbf{a}$) & 0.49(1)              & 10.48(1)               \\
	\end{tabular}
	\end{ruledtabular}
\caption{\label{phih} Parameters of power-law fits to the calorimetric phase boundary data in a temperature range $T<T_\mathrm{max}=0.6$~K. The analysis for \IPA in $\mathbf{H}\bot\mathbf{b}^\ast$ was based on data from Ref.~\onlinecite{Tsujii2009}.}
\end{table}

\subsection{Neutron diffraction}
One of the more spectacular observations of the previous neutron study of \IPAXP in the transverse field geometry was the substantial and history-dependent broadening of the antiferromagnetic Bragg peaks in the high field phase.\cite{Hong2010PRBRC} A key result of the present work is that this effect, while still present, is drastically reduced in the almost axially symmetric configuration.

\begin{figure}[htbp]
	\includegraphics[width=\columnwidth]{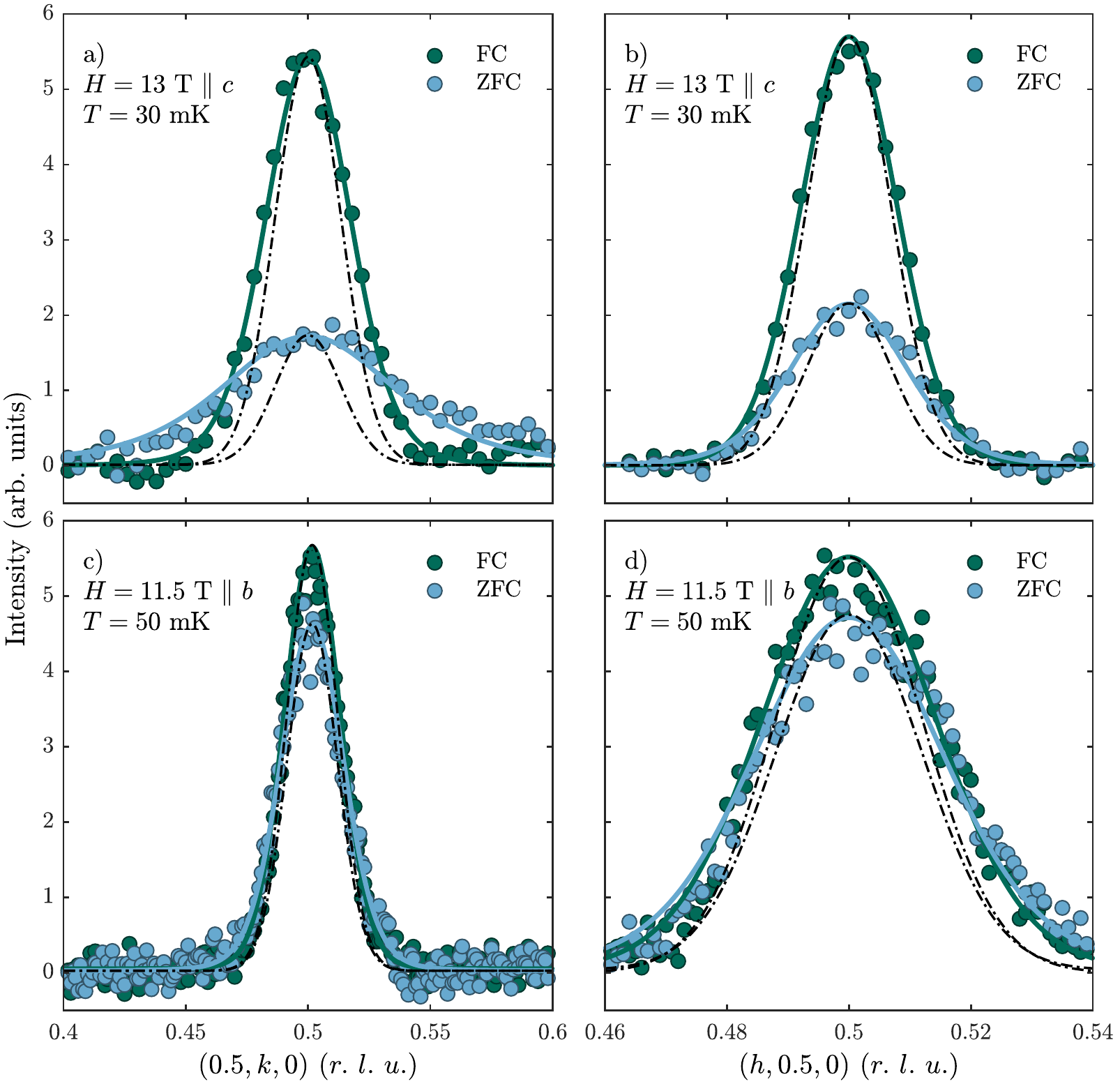}
	\caption{\label{scans} Magnetic $(0.5,0.5,0)$ Bragg peaks measured in \IPAXP in for a magnetic field $H=13$~T applied almost perpendicular (a, b) and and almost parallel  (c, d) to the easy anisotropy axis. The data were collected using field-colling (FC, open symbols) and zero-field-cooling (ZFC, closed symbols) protocols.
The dash-dot lines represent experimental Gaussian resolution that was measured as described in the text. Solid lines are fits to the data of Lorentzian-squared functions folded with the experimental resolution. The experimental data in panels (a) and (b) are from Ref.~\onlinecite{Hong2010PRBRC}.}
\end{figure}

Figure~\ref{scans} shows scans across the $(0.5,0.5,0)$ magnetic reflections in the high-field phase of \IPAXP for the magnetic field applied almost perpendicular (Ref.~\onlinecite{Hong2010PRBRC}) and almost parallel to the anisotropy axis (this work). These data were collected following either a field-cooling (FC) or zero-field cooling (ZFC) protocols. A Gaussian fit to instrumental resolution is in all cases shown in a dashed line. The resolution was determined by measuring the weak nuclear scattering contribution at $(0.5,0.5,0)$ due to $\lambda/2$ beam contamination. One immediately sees that the broadening and associated reduction in peak intensity of the ZFC peaks compared to FC ones is considerably smaller in the close to axial setting. In fact, in that case the peaks are almost resolution limited regardless of cooling protocol.

To quantify this observation we analyzed all the measured scans using Lorentzian-squared profiles\footnote{In Ref.~\onlinecite{Hong2010PRBRC} the correlation length was extracted in Lorentzian fits. However, using a Lorentzian-squared instead is more justified. It has the simple physical meaning of being the Fourier transform of an exponential correlation function in real space. Also, unlike that of the Lorentzian, its integral over reciprocal space is finite.} numerically convoluted with the instrument resolution. The fits are show in Figure~\ref{scans} in solid lines and allow us to extract the correlation length $\xi$, tabulated in Table~\ref{xi}.

\begin{table}[!htpb]
	\begin{ruledtabular}
	\begin{tabular}{lcc}
		    & $\xi_b$ (\AA) & $\xi_c$  (\AA)  \\ \hline
$\bot$-geometry (Ref.~\onlinecite{Hong2010PRBRC}):\\
		FC  & 30(4)              & 94(7)                \\
		ZFC & 8.3(1)              & 50(5)              \\ \hline

$\|$-geometry:\\
		FC  & 97(6)              & 83(4)              \\
		ZFC & 97(6)              & 83(4)               \\
	\end{tabular}
	\end{ruledtabular}
\caption{\label{xi} Magnetic correlation length measured in \IPAXP for a magnetic field applied almost perpendicular\cite{Hong2010PRBRC} and almost parallel to the anisotropy axis, in field-cooled and zero-field-cooled samples. The values are obtained from Lorentzian-squared fits to the diffraction data, as described in the text.}
\end{table}

\begin{figure}[htbp]
	\includegraphics[width=\columnwidth]{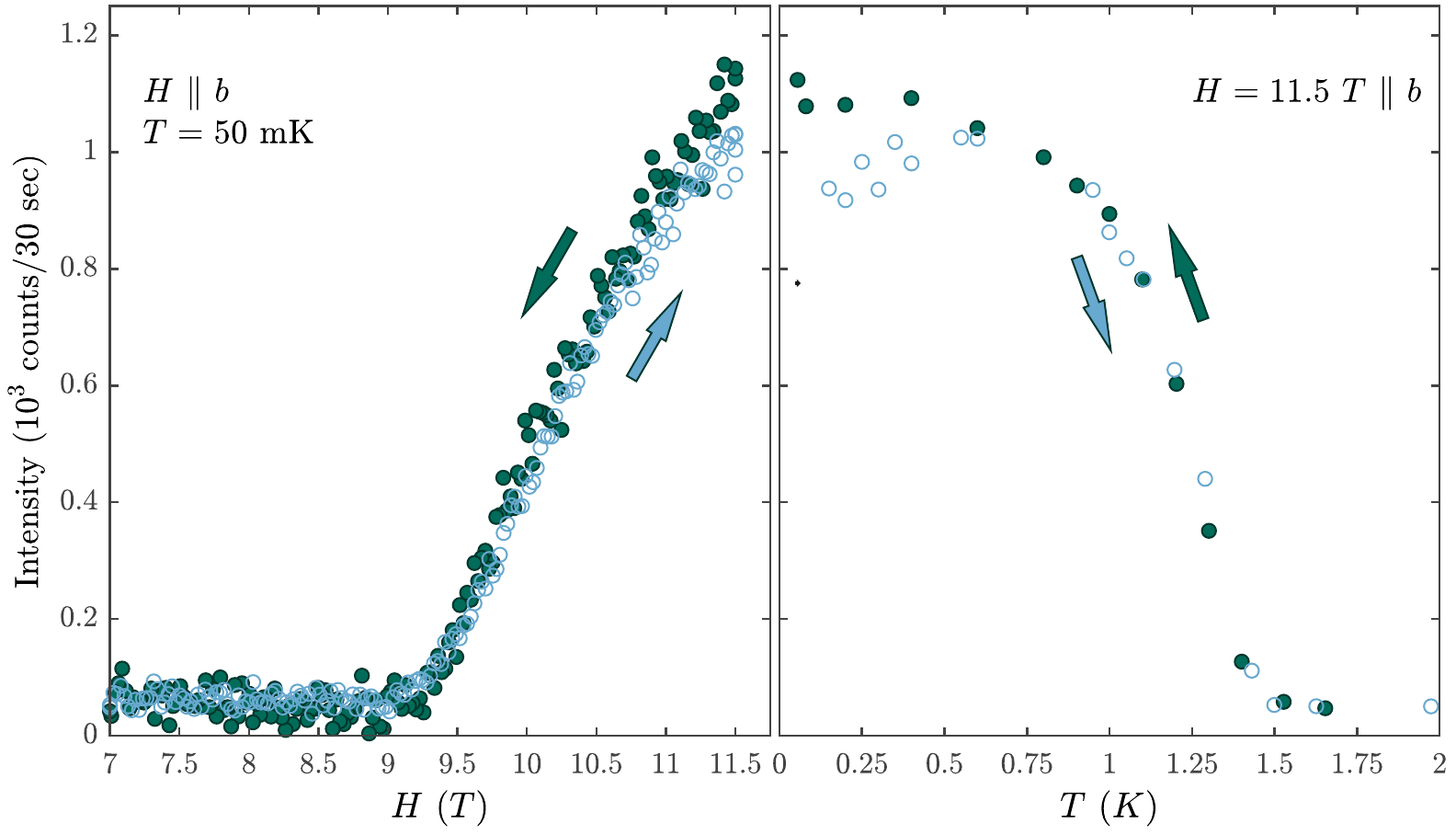}
	\caption{\label{hyst} Peak intensity of the $(0.5,0.5,0)$ Bragg reflection measured in \IPAXP for a magnetic field applied almost parallel to the easy anisotropy axis. The measurements are performed following zero-field-cooling the sample, and through an isothermal field ramping at low temperature, warming to above the transition tempereture, field-cooling and an isothermal ramping down of the applied field. The observed history dependence is much smaller than previously seen in the transverse-field geometry.~\cite{Hong2010PRBRC}}
\end{figure}

For a transverse field, the history-dependent short range order in  \IPAXP is also manifest in a hysteretic field- and temperature-dependencies of the diffraction peak height.\cite{Hong2010PRBRC} Such a hysteresis measurement for the close to axial geometry is shown in Fig.~\ref{hyst}. The sample was first cooled in zero field to $T=50$~mK, then the magnetic field was raised to $H=11.5$~T. Following that, the sample was warmed up to above the ordering temperature at a constant field, and the trajectory was reversed. While the hysteresis is clearly observed, it is considerable smaller than for a transverse field orientation.\cite{Hong2010PRBRC} Overall, in the observed behavior in the  close to axial geometry is similar to that previously seen in \PHCX, but with an even smaller hysteretic effect.\cite{Huevonen2012}

\section{Discussion}
The first thing to note is the clear differences in behavior of the disorder-free system in the two orientations. In agreement with Ref.~\onlinecite{Zhao2015}, for the off-axial field, BEC of magnons is a poor description for the criticality of the transition. The measured value of the phase boundary exponent is in better agreement with that of the 3+1 dimensional Ising model ($\phi=1/2$, since $z=1$ and $\nu=1/2$ at the upper critical dimension) then with $\phi=2/3$ expected for a 3+2 dimensional BEC of magnons (mean field).\cite{Giamarchi1999} That  the material behaves as an Ising system for $\mathbf{H}\bot\mathbf{b}$ at $T\lesssim 1$~K is, in retrospect, not surprising, since the anisotropy in the $(a^\ast,b^\ast)$-plane is itself of the order of 0.5~K.\cite{Manaka2001,Nafradi2011} Similar residual anisotropy has been a complication for most examples of magnon BEC in quantum magnets,\cite{Amore2009} including \TCCL.\cite{Glazkov2004,Kolezhuk2004} Moreover, in \IPA, due to a low crystal symmetry, a truly axially symmetric geometry can not be perfectly restored by an choice of field direction. Nevertheless, applying a field close to the principal axis of anisotropy seems to approximate the BEC scenario reasonably well, at least as far as the measured critical exponent is concerned.

For the disordered material, the short range ordering and history dependence is clearly strongest in the transverse field geometry. Therefore, it must be interpreted as an effect of disorder not on a BEC of magnons, but on a transition that is in the Ising universality  class. The chemical Cl/Br disorder in \IPAX will have several consequences for the magnetic Hamiltonian. First, the strength of antiferromagneic interactions will be affected due to variances in Cu-Cl and Cu-Br covalency strengths and distortions of superexchange bond angles. This type of disorder is exactly what translates into a random potential for bosons, as envisioned in the magnetic Bose Glass picture.\cite{Zheludev2013} In addition however, distortions of the local Cu$^{2+}$ crystallographic environments will lead ro a random component to the magnetic ion's gyromagnetic tensor. In the presence of a uniform external magnetic field, this will effectively generate a {\em random spin field} in the sample. It's component along the applied uniform field couples to magnetization, i.e., magnon density. Thus it adds to the random potential for magnons. However, the transverse component of the spin field is directly coupled to the order parameter of the phase transition, namely the transverse magnetization. Such random fields are known to have drastic effects on phase transitions.\cite{Imry1975} The third effect of chemical disorder is random two-ion (exchange) anisotropy. In the presence of a uniform external field a random anisotropy becomes again equivalent to a random field, with all the same consequences.\cite{Fishman1979} We conclude that for a field applied perpendicular to the principal anisotropy axis the phase transition in \IPAX is to be viewed as that in a Random Field and Random Bond (RF+RB) Ising transition, and surely not as a BEC to BG transition, as implied in Ref.~\onlinecite{Hong2010PRBRC}.

The ideal RF Ising model is known to order in three dimensions.\cite{Imry1975} The quantum and thermodynamic phase transitions are, in fact, governed by the same fluctuationless fixed-point.\cite{Senthil1998} Nevertheless, the actual behavior observed in prototype materials such as diluted AFs in a uniform field is quite different.\cite{Birgeneau1985,Birgeneau1983} Rather than showing long range order, these systems, of which Mn$_x$Zn$_{1-x}$F$_6$ is perhaps the best known example,\cite{Birgeneau1985}  go through a freezing transition. Below that point there is static but only short range order. The system breaks up into microscopic domains. The domain size is history dependent, but can only increase as long as one doesn't exit into the paramagnetic phase. There is also no temporal relaxation of domain size. This anomalous behavior has been attributed to the the simultaneous presence of  RF and RB.\cite{Nattermann1988} This situation exactly corresponds to \IPAX. Indeed, in the transverse field configuration we observe almost exactly the type of behavior as seen in Mn$_x$Zn$_{1-x}$F$_6$ at a fixed random-field strength.

The pinning of domains is governed by the width of the domain walls, i.e., by the relative magnitude of anisotropic vs. axially symmetric magnetic interactions.\cite{Nattermann1988} For \IPAX, the off-axial component of anisotropy is weakest in the close to axial geometries. In these configurations the pinning and resulting history-dependent behavior can thus expected to be much weaker than in the transverse-field case. This is totally consistent with our observations. In \IPAXP in a field applied almost along the anisotropy axis, the small hysteretic effects and internal line width aside, we observe what almost is a long-range-ordered phase. Perhaps by coincidence, adding disorder has almost no effect on the critical field $H_c$ in this case. Nonetheless, the phase boundary significantly modified. The measured critical exponent is very similar to that observed under similar conditions in \PHCX\cite{Huevonen2012} and \DTNX.\cite{Yu2012} Regardless of whether $\phi\sim 1$ is indeed the correct critical exponent for the BG to BEC transition or not,\cite{Yao2014} and whether that scenario is at all relevant for \IPAXP in the close-to-axial configuration, experimentally it seems to be a rather common situation in disordered magnets. Interestingly, for a transverse-field configuration the phase boundary critical exponent appears to be unchanged by the presence of disorder.

In summary, random anisotropy and random field effects can never be fully disregarded in field-induced phase transitions in materials with chemical disorder, even if the latter does not involve the magnetic ions directly. This work was partially supported by the Swiss National Science foundation, Division II.

\end{document}